\documentclass[12pt]{article}
\usepackage{amsfonts,amsmath,amssymb}

\title{Are All Particles Identical?}
\author{
Sheldon Goldstein\footnote{Departments of Mathematics, Physics and
       Philosophy, Hill Center, Rutgers, The State University of New
       Jersey, 110 Frelinghuysen Road, Piscataway, NJ 08854-8019, USA.
       E-mail: oldstein@math.rutgers.edu},
James Taylor\footnote{Department of Mathematics, Iowa State
       University, Carver Hall, Ames, IA 50010, USA. E-mail:
       jostylr@member.ams.org},\\
Roderich Tumulka\footnote{Dipartimento di Fisica dell'Universit\`a di
       Genova and INFN sezione di Genova, Via Dodecaneso 33, 16146
       Genova, Italy. E-mail: tumulka@mathematik.uni-muenchen.de},\
       and
Nino Zangh\`\i\footnote{Dipartimento di Fisica dell'Universit\`a di
       Genova and INFN sezione di Genova, Via Dodecaneso 33, 16146
       Genova, Italy. E-mail: zanghi@ge.infn.it}
}
\date{September 28, 2004}

\newcommand{\RdN}{\mathbb{R}^{3N}}

\newcommand{\NRd}{{}^{N}\mathbb{R}^3}
\newcommand{\Rd}{\mathbb{R}^3}
\newcommand{\RRR}{\mathbb{R}}
\newcommand{\CCC}{\mathbb{C}}
\newcommand{\vx}{\boldsymbol{x}}
\newcommand{\vy}{\boldsymbol{y}}
\newcommand{\vq}{{\boldsymbol{q}}}
\newcommand{\vQ}{{\boldsymbol{Q}}}
\newcommand{\vj}{{\boldsymbol{j}}}
\newcommand{\vv}{{\boldsymbol{v}}}
\newcommand{\I}{i}

\newcommand{\Q}{\mathcal{Q}}

\renewcommand{\Im}{\mathrm{Im}}
\newcommand{\valuespace}{\mathbb{W}}

\begin{document}
\maketitle
\begin{abstract}
   We consider the possibility that all particles in the world are
   fundamentally identical, i.e., belong to the same species.
   Different masses, charges, spins, flavors, or colors then merely
   correspond to different quantum states of the same particle, just as
   spin-up and spin-down do. The implications of this viewpoint can be
   best appreciated within Bohmian mechanics, a precise formulation of
   quantum mechanics with particle trajectories.  The implementation of
   this viewpoint in such a theory leads to trajectories different
   {}from those of the usual formulation, and thus to a version of
   Bohmian mechanics that is inequivalent to, though arguably
   empirically indistinguishable {}from, the usual one. The
   mathematical core of this viewpoint is however rather independent of
   the detailed dynamical scheme Bohmian mechanics provides, and it
   amounts to the assertion that the configuration space for $N$
   particles, even $N$ ``distinguishable particles,'' is the set of all
   $N$-point subsets of physical 3-space.

   \medskip

\end{abstract}

\section{Introduction and Overview}\label{sec:io}

It is not a new idea that what appear to be two different species of
particles may in fact be two different states of the same species.  It
is particularly obvious that spin-up and spin-down are merely two
states of the same particle because we often encounter superpositions,
such as spin-left, of these states. But even in cases in which
superpositions are extremely hard to obtain, such as of different
quark flavors, it is not an unusual idea that what is behind is the
same particle, the quark.  One may also consider the thought that
electrons and positrons are, despite their difference in charge, the
same particle; this is suggested in particular by the fact that, while
electron states are positive energy solutions of the Dirac equation,
positron states are conjugate negative energy solutions of the Dirac
equation---a situation similar to left-handed and right-handed
photons.  Moreover, any explanation of the particular values of masses
of the elementary particles would probably have to consist in deriving
the appropriate energy eigenvalues for different states---of the same
particle.  Note also that supersymmetry suggests that the same
particle can appear as a boson or a fermion, another example of how
one species can appear as two.

We explore here the most extreme possibility of this kind: that all
particles are fundamentally identical, i.e., that fundamentally only
one species of elementary particles exists.  Let us call this the {\em
   identity hypothesis}.  This one species would then have to have
different states corresponding to being an electron, a quark, a
neutrino, or whatever.  The consequences of the identity hypothesis
are, in a sense, less dramatic than one might expect: we shall point
out how \emph{every} quantum theory involving several particle species
can indeed be transformed into a theory of just one species, thus
incorporating the identity hypothesis, \emph{without any change in the
   predictions for experiments}.

The identity hypothesis can be paraphrased as a statement about the
configuration space.  The configuration space of a universe of $N$
distinct particles is $\RdN$, whereas the configuration space of a
universe of $N$ identical particles is, as argued in \cite{LDW71,
Leinaas, stochmech, BSB99, identical}, the set of all $N$-element
subsets of $\Rd$,
\begin{equation}
\label{NR3def}
    \NRd := \{ S \subseteq \Rd | \# S = N \}\,.
\end{equation}
This is a manifold of dimension $3N$ (for a mathematical discussion of
this manifold, see \cite{identical,Leinaas}). While a configuration in
$\RdN$ indicates that particle 1 is at location $\vQ_1 \in \Rd$, etc.,
a configuration {}from $\NRd$ provides just $N$ points in $\Rd$, but
no further information on which particle is where.  The identity
hypothesis amounts to the statement that $\NRd$ should {\em always} be
considered the natural configuration space, even for a system of $N$
``distinguishable'' particles. As we shall show below, an
implementation of this idea is provided by a unitary transformation
{}from the standard representation of wave functions as elements of
$L^2(\RdN, \CCC^{k_1} \otimes \cdots \otimes \CCC^{k_N})$, to a
new representation of square-integrable cross-sections of a suitable
bundle based on $\NRd$.

To appreciate our proposal, it is helpful to keep in mind that even if
a theory treats particles of different species as different sorts of
points, associated with different masses, etc., the associated
properties are not directly accessible to an observer. When we want to
find out whether a given particle is, say, an electron or a muon, we
study the particle's reaction to various electric fields or other
external conditions under our control. For example, when a wavepacket
that is a superposition of an electron and a muon is subjected to an
electric field, the two contributions to the packet diverge due to
their difference in the charge--mass ratio, thus forming two disjoint
packets; when the particle is finally detected in the muon packet, we
will say it is a muon. Thus, we know that a point in front of us is a
muon point rather than an electron point, not because we look into its
essence---whatever that would mean---but because its nature is encoded
in patterns in its environment. Once one realizes this, the identity
hypothesis, which might at first appear to be an outrageous notion,
seems like a real possibility.

Within Bohmian mechanics|a precise formulation of quantum mechanics
accounting for all quantum phenomena in terms of point particles
moving in physical space \cite{Bohm52, DGZ, Stanford,op}|this
hypothesis acquires an even stronger justification since the theory is
{\em primarily} about particles, with the wave function having the
dynamical role of governing their motion. The choice of $\NRd$ as
configuration space corresponds to the insistence that the actual
configuration of an $N$-particle system be a set of $N$ points in
physical space, with the points labeled in no way, neither by numbers
$1, \ldots, N$, nor in the sense that there could be different kinds
of points in the world, such as electron points as distinct {}from
muon points or quark points.  Given merely an actual configuration set
$Q\in \NRd$, there is then simply no fact in the world about
\emph{what sort} of particle there is at a point $\vQ\in Q $ in
physical space, only \emph{that} there is a particle.  This particle
is not associated with any label, mass, charge, spin, flavor, or
color.

As we shall explain, the identity hypothesis has sharp consequences in
the context of Bohmian mechanics: the law of motion has to be
suitably adjusted, thus leading to trajectories different {}from those
of the conventional version of the theory. Therefore, the identity
hypothesis has a genuine, nontrivial meaning in Bohmian mechanics (and
as well in stochastic mechanics \cite{stochmech}).  However, arguably,
no possible experiment can confirm or disprove the identity hypothesis
in the sense of distinguishing between these two versions of Bohmian
mechanics.  Hence, the identity question remains empirically
undecidable---unless one day progress in physics leads to a refined,
empirically testable theory of elementary particles that needs the
identity hypothesis.  However, empirically undecidable does not mean
completely undecidable. There may be relevant differences in
simplicity and naturalness.

To our knowledge, the consequences of the identity hypothesis have
never been discussed in the literature. We are not the first, however,
to consider the identity hypothesis.  It is implicit in Bell's
``Beables for quantum field theory'' \cite{BellBeables}, which has
inspired this work.  Bell's model, like Bohmian mechanics, involves
additional variables beyond the wave function, and as in Bohmian
mechanics, these variables are basically particle positions.  More
precisely, a configuration in Bell's model is described by specifying
the number of particles present at every location in (a discrete
version of) 3-space.  These values are regarded as the actual values
of a certain family of observables, given by the fermion number
operator at each point in 3-space, which is the sum of the number
operators over all (fermion) particle species. Thus, a configuration
in Bell's model, like one in $\NRd$, does not distinguish between
different kinds of particles.

\section{Standard Bohmian Mechanics}\label{sec:2vbm}

We begin by describing the conventional version of Bohmian mechanics
\cite{Bohm52,Stanford}.  For the sake of concreteness, let $\psi: \RdN
\to \CCC^{k_1} \otimes \cdots \otimes \CCC^{k_N} = \valuespace$ be a
quantum mechanical wave function of an $N$-particle system, obeying
(as a low-energy description) the Schr\"odinger equation,
\begin{equation}
\label{Schroedinger}
    \I\hbar \frac{ \partial \psi}{ \partial t} = - \sum_{i=1}^N
    \frac{\hbar^2}{2m_i} \nabla_i \cdot \nabla_i \psi + V\psi\,,
\end{equation}
where $m_i$ denotes the mass of the $i$-th particle, and $V$ the
potential which is possibly (Hermitian) matrix valued. The several
components of $\psi \in \valuespace$ represent internal degrees of
freedom such as spin, flavor, or color.  In the usual version of
Bohmian mechanics, the $i$-th particle moves according to
\begin{equation}
\label{Bohm}
    \frac{d\vQ_i}{dt} = \frac{\vj_i(Q)}{\rho(Q)}
\end{equation}
where $Q = (\vQ_1, \ldots, \vQ_N)$ is the configuration in
$\RRR^{3N}$,
\begin{equation}
\label{rhoRdN}
    \rho = \psi^* \psi
\end{equation}
(meaning a scalar product in $\valuespace$) is the probability density,
and
\begin{equation}
\label{jidef}
    \vj_i = \frac{\hbar}{m_i}\, \Im \, \psi^* \nabla_i \psi
\end{equation}
is the probability current, more precisely the part of it
corresponding to the $i$-th particle.

We have that if $Q(0)$ is random with distribution $\rho_0$ (as we
shall assume in the following), then $Q(t)$ has distribution $\rho_t$
for every time $t$.  This follows {}from the fact that $\rho$ and the
velocity field $v$ on $\RRR^{3N}$ defined in \eqref{Bohm} and
\eqref{rhoRdN} obey, by virtue of the Schr\"odinger equation
\eqref{Schroedinger}, a continuity equation
\begin{equation}
\label{continuity}
    \frac{
\partial \rho}{
\partial t} = -\mathrm{div} (\rho v)
\end{equation}
on $\RRR^{3N}$.  This is an extremely important property, as it
expresses a certain compatibility between the two equations of motion
defining the dynamics, which we call the \emph{equivariance} of
$\rho=|\psi|^2$ \cite{DGZ}.  Such a notion plays a crucial role in
establishing the empirical import of the theory \cite{Bohm52, DGZ,
   op}.

\section{Identity-based Bohmian Mechanics}\label{sec:v}

Bohmian versions of quantum theory define a dynamics on some
configuration space $\Q$; the choice $\Q = \NRd$ reflects the identity
hypothesis, as the particles themselves are then not associated with
labels, masses, charges, etc.---which are implicit, though, in the
wave function, and relevant to its time evolution
\eqref{Schroedinger}. With such a choice the actual configuration
$Q(t)$ would move in $\NRd$, whereas $\RdN$ would play a role merely
for the wave function.

A crucial observation is that a dynamics on $\NRd$ is given by a
symmetric (permutation invariant) dynamics on $\RdN$. However,
\eqref{Bohm} is not symmetric (and thus does not define a dynamics on
$\NRd$ and fails to be compatible with the identity hypothesis),
except in the special case that the masses are all the same, the
potential $V$ is permutation invariant, and the wave function is
symmetric or antisymmetric.

We are thus led to consider the dynamics on $\RdN$ obtained by
replacing \eqref{Bohm} {}from the conventional version with
\begin{equation}
\label{Bohmsym}
    \frac{d\vQ_i}{dt} = \frac{\sum_\sigma \vj_{\sigma(i)} (\sigma Q)}
    {\sum_\sigma \rho(\sigma Q)}
\end{equation}
where the sums are taken over the group $S_N$ of permutations of $N$
elements, and $\sigma Q = (\vQ_{\sigma^{-1}(1)}, \ldots,
\vQ_{\sigma^{-1}(N)})$.\footnote{This means the particle at $\vQ_i$
   gets the number $\sigma(i)$, defining its new place in the ordering.
   Note that $\vj_{\sigma(i)}(\sigma Q)$ lies, for every $\sigma$, in
   the tangent space to $\Rd$ at $\vQ_i$, if we wish to distinguish
   between tangent spaces at different points.  } In words, the
velocity of \eqref{Bohmsym} is obtained {}from the one of \eqref{Bohm}
by \emph{symmetrizing} the current and the density, i.e., by averaging
over all permutations of the $N$ particles. Note that this is
different {}from symmetrizing the wave function, a procedure that
would lead to zero in the presence of fermions. Note also that this is
different {}from directly symmetrizing the velocity \eqref{Bohm},
which would also define a dynamics on $\NRd$, but one for which,
unlike with \eqref{Bohmsym}, the dynamics obtained will not be
equivariant, see below. Since \eqref{Bohm} and \eqref{Bohmsym} define
different velocities, they lead to different trajectories and thus
inequivalent theories.

We emphasize that the theory defined by \eqref{Bohmsym} is permutation
invariant in a strong sense, a sense in which the one defined by
\eqref{Bohm} is not: if $Q(t)$ and $Q'(t)$ are two solution curves of
\eqref{Bohmsym} in configuration space $\RdN$, and if at some time $Q' =
\sigma Q$ for some permutation $\sigma \in S_N$, then $Q'(t) = \sigma Q(t)$
at all times $t$.  By virtue of this permutation invariance,
\eqref{Bohmsym} defines a dynamics on $\NRd$ via projection. After all,
if we know the set of $N$ points contained in the configuration $Q \in
\RdN$ at some time, then this information is sufficient to determine the
configuration $Q(t)$ at all times, modulo the ordering.  We call the
dynamics on $\NRd$ thus obtained {\em identity-based Bohmian mechanics}.

To get a handle on identity-based Bohmian mechanics and its symmetry
properties, consider the very simple example of two particles, say an
electron (particle 1) and a muon (particle 2), with scalar-valued wave
function $\psi (q_{1}, q_{2})= \phi(q_{1}) \chi(q_{2})$, which are,
say, evolving according to the free Hamiltonian (so that factorization
is preserved). Then, according to \eqref{Bohm}, the equations of
motion of standard Bohmian mechanics are
    \begin{eqnarray}
    \frac{d\vQ_{1}}{dt} &=& \frac{\hbar}{m_{e}} \Im \, \frac{ \nabla \phi
(\vQ_{1})}{\phi (\vQ_{1})}  \label{b1}\\
     \frac{d\vQ_{2}}{dt} &=& \frac{\hbar}{m_{\mu}} \Im \, \frac{
\nabla\chi (\vQ_{2})}{\chi (\vQ_{2})}\label{b2}
   \end{eqnarray}
   where $m_{e}$ and $m_{\mu}$ are respectively the masses of the
   electron and of the muon. On the other hand, according to
   \eqref{Bohmsym}, the dynamics of motion of identity-based Bohmian
   mechanics is given by
    \begin{eqnarray}
    \frac{d\vQ_{1}}{dt} &=& \frac{   \frac{\hbar}{m_{e}}
|\chi(\vQ_{2})|^{2} \Im\, [ \phi(\vQ_{1})^{*} (\nabla \phi)(\vQ_{1})  ]
+  \frac{\hbar}{m_{\mu}}  |\phi(\vQ_{2})|^{2} \Im\, [ \chi(\vQ_{1})^{*}
(\nabla \chi)(\vQ_{1})  ] }{|\phi(\vQ_{1})|^{2}  |\chi(\vQ_{2})|^{2} +
|\phi(\vQ_{2})|^{2}  |\chi(\vQ_{1})|^{2} } \label{bs1}\\
     \frac{d\vQ_{2}}{dt} &=& \frac{   \frac{\hbar}{m_{\mu}}
|\phi(\vQ_{1})|^{2} \Im\, [ \chi(\vQ_{2})^{*} (\nabla \chi)(\vQ_{2})  ]
+  \frac{\hbar}{m_{e}}  |\chi(\vQ_{1})|^{2} \Im\, [ \phi(\vQ_{2})^{*}
(\nabla \phi)(\vQ_{2})  ] }{|\phi(\vQ_{1})|^{2}  |\chi(\vQ_{2})|^{2} +
|\phi(\vQ_{2})|^{2}  |\chi(\vQ_{1})|^{2}} \label{bs2}
     \end{eqnarray}

     Note that now the two particle indices 1 and 2 do not carry any
     direct relation to the particle species (as characterized by their
     masses): Contrary to the velocity formulas \eqref{b1} and
     \eqref{b2}, there is nothing in the right hand side of \eqref{bs1}
     and \eqref{bs2} that distinguishes $\vQ_{1}$ {}from $\vQ_{2}$. The
     two particles are distinguished, rather, {\em only} by the
     positions they happen to have at any given time. Note in
     particular that if $\phi$ and $\chi$ have disjoint supports, with,
     say, $\phi$ supported on ``the left'' and $\chi$ on ``the right,''
     then only one term in the numerator and one in the denominator of
     \eqref{bs1} and \eqref{bs2} will be nonvanishing.  Then $\vQ_{1}$
     will behave like an electron, resp.\ like a muon, when it is on the
     left, resp. right. In other words, the particle on the left
     behaves like an electron and the one on the right like a muon,
     regardless of how we might label them.

     More generally, for arbitrary two-particle wave function and
     Hamiltonian, consider a configuration with an electron at $\vx \in
     \Rd$ and a muon at $\vy \in \Rd$, and another configuration with
     the muon at $\vx$ and the electron at $\vy$. In both
     configurations, \eqref{Bohmsym} yields the same velocity for the
     particle at $\vx$, regardless of whether it is the electron or the
     muon (and the same velocity at $\vy$). With \eqref{Bohm}, in
     contrast, there is no reason why this should be so.

     \bigskip

The equation of motion for $Q\in\NRd$ can be written as
\begin{equation}
\label{dynamics}
    \frac{dQ}{dt} =v_t(Q)
\end{equation}
where $v=v_t$ is a time-dependent vector field on $\NRd$ that is
determined by \eqref{Bohmsym}.  Similarly, the denominator of
\eqref{Bohmsym} defines a probability density $\rho$ on $\NRd$, while
its numerator defines a probability current $J=\rho v$ on $\NRd$.

As in conventional Bohmian mechanics, also in identity-based Bohmian
mechanics we have that if $Q(0)$ is chosen at random with distribution
$\rho_0$, then $Q(t)$ has distribution $\rho_t$ for every time $t$.
This is a consequence of the fact that $\rho$ and $v$ obey the
continuity equation
\begin{equation}
\label{econtinuity}
    \frac{
\partial \rho}{
\partial t} = -\mathrm{div} (\rho v)
\end{equation}
on $\NRd$, which follows {}from \eqref{continuity} by summing both
sides over all permutations.

\section{Wave Function on $\NRd$}

If the configuration space is $\Q = \NRd$, then one might imagine that
the wave function should also live on $\NRd$ rather than on $\RdN$.
And that is possible! One may transform any wave function $\psi: \RdN
\to \valuespace = \CCC^{k_1} \otimes \cdots \otimes \CCC^{k_N}$ into a
cross-section $\phi$ of a suitable vector bundle $E$ over $\NRd$.  The
fiber space $E_q$ at $q \in \NRd$ of this bundle has dimension $N!k_1
\cdots k_N$ and can be defined as
\begin{equation}
\label{fiber}
    E_q = \bigoplus_{\nu\in B_q} \valuespace
\end{equation}
where $B_q$ is the set of all bijections $q \to \{ 1, \ldots, N\}$;
thus, $\nu$ runs through all possible numberings, or all possible
identifications of the $N$ points with the $N$ particle
``identities.'' In particular, $E_q$ is the direct sum of $N!$ copies
of $\valuespace$.  (It follows that even for $k_1= \cdots = k_N=1$
(spinless particles), the wave function $\phi(q)$ has $N!$ components
at $q \in \Q$, each component corresponding to a particular way of
labeling the points.) We define the scalar product in $E_q$ in such a
way that \eqref{fiber} is an \emph{orthogonal} sum.

Another way of viewing $E_q$ is this: every $\nu \in B_q$ defines an
ordering of the $N$ points of $q$, and thus an element of $\RdN$,
namely $\hat{q} = (\nu^{-1}(1), \ldots, \nu^{-1}(N))$. We shall
sometimes write $(q,\nu)$ for $\hat{q}$. $E_q$ is the direct sum of
the value spaces (each being a copy of $\valuespace$) of $\psi$ at the
points $\hat{q} \in \RdN$. (This view of $E_q$ allows a generalization
to the case that $\psi$ is itself a cross-section of a vector bundle
over $\RdN$.) And this view is, in fact, how the transformation of
$\psi$
into $\phi$ can best be understood:
\begin{equation}
\label{phipsi}
    \phi(q) = \oplus_\nu \psi(q,\nu) = \oplus_\nu \psi(\hat{q}).
\end{equation}
Conversely, $\psi$ can be reconstructed {}from $\phi$ at all
configurations $(\vq_1, \ldots, \vq_N)$ without coincidences (i.e.,
such that $\vq_i \neq \vq_j$ whenever $i\neq j$):
\begin{equation}
\label{psiphi}
    \psi(\vq_1, \ldots, \vq_N) = \phi_\nu (\{\vq_1, \ldots, \vq_N\})
\end{equation}
with $\nu(\vq_i) :=i$.  Since the configurations with coincidences
form a null set in $\RdN$, the transformation $\psi \mapsto \phi$
defines a unitary identification between the Hilbert spaces $L^2(\RdN,
\valuespace) \to L^2(E)$, where by $L^2(E)$ we denote the space of
square-integrable cross-sections of the bundle $E$.  \bigskip

(We note in passing that in identity-based Bohmian mechanics the
notion of {\em conditional wave function} for subsystems \cite{DGZ} is
defined much less often than in standard Bohmian mechanics. For
example, in standard Bohmian mechanics the conditional wave function
for scalar-valued wave functions is always defined, but this is not
the case in identity-based Bohmian mechanics, due to the fact that the
``particle identities'' are then internal degrees of freedom (see
above), more or less like spin in standard Bohmian mechanics. Thus in
identity-based Bohmian mechanics only the notion of conditional
density matrix is always well defined \cite{density}. This may appear
paradoxical, in view of our claim that the two theories are
empirically equivalent (see below). There is, however, no paradox:
upon reflection we realize that whenever we would {\em know}, in a
standard Bohmian universe, that the wave function of a subsystem is
$\psi$, then, for identity based Bohmian mechanics, its conditional
density matrix would in fact be $|\psi\rangle \langle \psi|$.)

\section{Dynamics in Terms of $\phi$}

For the sake of completeness, we explicitly describe in this section
how the identity-based Bohmian theory can be formulated purely in
terms of the bundle cross-section $\phi$.  First note that
\[
|\phi(q)|^2 = \sum_{\nu\in B_q} |\psi(q,\nu)|^2,
\]
so that $\rho(q) = |\phi(q)|^2$.  A connection (covariant derivative
operator) $\nabla$ can be defined on $E$ in an obvious
way.\footnote{Here is the definition, in terms of parallel transport:
   if $q(s)$, $0 \leq s \leq 1$, is a curve in $\NRd$ then any
   bijection $\nu_0: q(0) \to \{1,\ldots, N\}$ can be transported along
   the curve by having the numbering follow the points of $q(s)$ as
   they move continuously {}from $q(0)$ to $q(1)$, thus defining a
   final bijection $\nu_1: q(1) \to \{1,\ldots, N\}$. Parallel
   transport of some element $e(0) \in E_{q(0)}$ along the curve leads
   to $\oplus_{\nu_1} e_{\nu_0}(0) =: e(1) \in E_{q(1)}$. This
   connection is nontrivial (i.e., parallel transport along a loop may
   differ {}from the identity) but flat (i.e., has curvature zero).}
The component of the probability current relative to the particle at
$\vq \in q$ is given by
\[
\vj_\vq (q) = \sum_\nu \frac{\hbar}{m_{\nu(\vq)}} \, \Im \, \phi^*_\nu
(q) \nabla_{\vq} \phi_\nu(q)\,,
\]
and we obtain the following formula for the vector field $v$ in terms
of $\phi$:
\[
\vv_\vq = \frac{\vj_\vq}{\rho}\,,
\]
where $\vv_\vq$ is the component of $v(q)$ corresponding to $\vq \in
q$.  In another notation,
\begin{equation}
\label{BohmsymQ}
    \frac{d\vQ}{dt} =\sum_\nu \frac{\hbar}{m_{\nu(\vQ)}} \, \Im
    \, \frac{ \phi^*_\nu (Q) \nabla_{\! \vQ} \phi_\nu(Q)}{\phi^* (Q) \,
    \phi(Q)}\,.
\end{equation}
The time evolution of $\phi$ obeys
\begin{equation}
\label{phiSchroedinger}
    \I \hbar \frac{
\partial \phi_\nu}{
\partial t}(q) = - \sum_{\vq \in q}
    \frac{\hbar^2}{2m_{\nu(\vq)}} \nabla_\vq \cdot \nabla_\vq
\phi_\nu(q)
    + V(q,\nu) \phi_\nu(q)\,.
\end{equation}

\section{A Remark on Identical Particles}\label{sec:id}

It is interesting to see what happens when we apply the transformation
\eqref{phipsi} to a system of identical particles, i.e., of bosons or
fermions. If $\psi$, with value space $\valuespace = (\CCC^k)^{\otimes
N}$, represents a system of $N$ bosons, then it must be symmetric
under permutation. As a consequence, $\psi(q,\nu) = R_{\nu' \circ
\nu^{-1}} \psi(q,\nu')$, where for any permutation $\sigma \in S_N$ of
the particles, $R_{\sigma}: (\CCC^k)^{\otimes N}\to (\CCC^k)^{\otimes
N}$ is the linear mapping that correspondingly permutes the components
of $\psi$. It follows that
\begin{equation}
\label{phiRphi}
    \phi_\nu = R_{\nu' \circ \nu^{-1}} \phi_{\nu'}\,,
\end{equation}
so that $\phi$ is actually confined to a subbundle of $E$,
characterized by \eqref{phiRphi} and having dimension $k^{N}$ rather
than $N!k^{N}$. The subbundle is parallel with respect to the
connection, i.e., parallel transport will remain within the subbundle.
For $k=1$, $R_\sigma$ is the identity, and the subbundle defined by
\eqref{phiRphi} is the trivial bundle $\Q \times \CCC$, so that $\phi$
can be identified with a function $\NRd \to \CCC$.

Now consider a system of fermions. Then $\psi$ is anti-symmetric under
permutations, so that $\psi(q,\nu) = (-1)^{\nu' \circ \nu^{-1}}
R_{\nu' \circ \nu^{-1}} \psi(q,\nu')$ where $(-1)^\sigma$ denotes the
sign of the permutation $\sigma$, and consequently,
\begin{equation}
\label{phiminusRphi}
    \phi_\nu = (-1)^{\nu' \circ \nu^{-1}} R_{\nu' \circ \nu^{-1}}
\phi_{\nu'}\,.
\end{equation}
In this case, $\phi$ is confined to another parallel subbundle of $E$,
characterized by \eqref{phiminusRphi} and also having dimension
$k^{N}$.  For $k=1$, this subbundle can be called the Fermi line
bundle; it has been described in \cite{Leinaas,Jamesthesis,identical}.
The connection is such that parallel transport along a closed
curve in $\NRd$ that realizes a permutation $\sigma$ coincides with
multiplication by $(-1)^\sigma$, the sign of $\sigma$.

Finally, when the system of $N$ particles under consideration contains
$N_1$ identical particles of species 1, \ldots, and $N_\ell$ identical
particles of species $\ell$, so that $N_1 + \ldots + N_\ell = N$, then
$\phi$ is confined to a subbundle $F$ of dimension $(N!/N_1!  \cdots
N_\ell!)\, k_{1}^{N_{1}}\cdots k_{\ell}^{N_{\ell}} $, characterized by
one condition like either \eqref{phiRphi} or \eqref{phiminusRphi} for
each species.

\section{Another Approach to Identity-Based Bohmian Mechanics}

In this section we briefly describe a different perspective on
identity-based Bohmian mechanics that lies outside the main line of
this paper and in particular is not essential for an understanding of
our main points.  We have, in the previous sections, formulated
identity-based Bohmian mechanics on $\NRd$, starting from conventional
Bohmian mechanics on $\RdN$. This brings us to the question as to
whether one could arrive at this formulation without ever invoking
$\RdN$.  We now sketch how to do this for bosons.  For a discussion of
fermions see \cite[Section~5.9]{Jamesthesis}.

Suppose that the state space $W$, representing the internal degrees of
freedom for a single particle, is the sum of the state spaces $W_j$ of
$\ell$ different species,
\[
  W = \bigoplus_{j=1}^\ell W_j, \quad W_j = \CCC^{k_j}.
\]
This means that the particle can be in a superposition of being an
electron, a muon, etc.  For $N$ bosons with state space $W$, the wave
function $\phi$ is a cross-section of the bundle $E'$ defined by
\[
  E'_q = \bigotimes_{\vq \in q} W;
\]
see \cite{identical} for a detailed discussion. The requirement that
we have $N_1$ particles of species 1, \ldots, and $N_\ell$ particles
of species $\ell$ defines a subbundle $F'$ of $E'$ which can be
identified with the bundle $F$ introduced in the last paragraph of
Section~\ref{sec:id}, in the case of bosons.  A cross-section of $F'$
remains in $F'$ under the Schr\"odinger evolution, and the obvious
Bohmian dynamics on $\NRd$ associated with cross-sections of $E'$ (see
\cite{Jamesthesis,identical} for the explicit definition) agrees for
cross-sections of $F'$ with the dynamics defined by \eqref{BohmsymQ}
for the corresponding cross-section of $F$.

\section{Empirical Equivalence}

One may wonder whether identity-based and conventional Bohmian mechanics
are empirically distinguishable, i.e., whether any possible
experiment could enable us to decide between these two versions of Bohmian
mechanics.  The question is delicate since it relates to the meaning of
empirical equivalence in general; we touch a bit more on this issue in a
separate work \cite{conf}. That the answer arguably is no can however be
appreciated quite easily by reflecting upon the three following points:
\begin{itemize}

\item The outcomes of all conceivable experiments will be recorded in
   the unordered configuration $\{\vQ_1, \ldots, \vQ_N\}$. To
   illustrate this fact, we may imagine the outcome as given by the
   orientation of a pointer on a scale; as the pointer consists of a
   huge number of electrons and quarks, for reading off the orientation
   of the pointer we need not be explicitly told which points are the
   electrons and which are the quarks.

\item The label of a particle in a standard Bohmian universe is not
   directly perceptible to an observer. We would base the decision
   whether, say, a given particle is an electron or a muon on how the
   particle moves under certain conditions that we control, say in
   terms of wave packets that spatially diverge due to differences (as
   the charge--mass ratio) encoded in the wave function\footnote{For
   an elementary illustration of this fact, recall the example
   discussed in Section \ref{sec:v}: as remarked there, when the
   supports of $\phi$ and $\chi$ are disjoint, \eqref{bs1} and
   \eqref{bs2} reduce to \eqref{b1} and \eqref{b2}, up to a possible
   permutation of the indices 1 and 2.} and finally grounded (after
   the experiment has been completed) in macroscopic patterns in the
   environment. This fact strongly suggests that only $\{\vQ_1,
   \ldots, \vQ_N\}$ and not $(\vQ_1, \ldots, \vQ_N)$ could ultimately
   be empirically relevant.

\item In conventional Bohmian mechanics, the distribution of the
   configuration $Q$ is $|\psi|^2$ at any time (see the end of Section
   \ref{sec:2vbm}).\footnote{In the case that empirical evidence
   contradicting quantum mechanics should be found, such as, e.g.,
   evidence for spontaneous wave function collapse \cite{BG03} or for
   violations of the quantum equilibrium distribution $|\psi|^2$ (as
   studied particularly by A.~Valentini \cite{Val02}), both
   conventional and identity-based Bohmian mechanics [understood as
   (quantum) equilibrium theories], though still empirically
   equivalent to each other, would become empirically inadequate:
   their predictions would be wrong.} If we ignore the labeling of the
   particles, we obtain {}from $Q= (\vQ_1, \ldots, \vQ_N)$ the set
   $\{\vQ_1, \ldots, \vQ_N\} \in \NRd$ whose (marginal) distribution
   coincides with the distribution $\rho$ of the configuration of
   identity-based Bohmian mechanics (see the end of Section
   \ref{sec:v}).  Since any empirical decision, if it can be made at
   some time (after, say, an experiment has been performed), must be
   based on the configuration (of systems, apparatuses, and the rest
   of the universe) at that time, two theories predicting the same
   distribution for the configuration $\{\vQ_1, \ldots, \vQ_N\}$
   cannot presumably be distinguished by experiment.

\end{itemize}

To sum up, it seems that no experiment could enable us to decide
between conventional and identity-based Bohmian mechanics. It is
presumably also true of orthodox quantum mechanics that no experiment
could enable us to distinguish an identity-based version from the
conventional version; this conclusion is based on the \emph{special
status of position measurements}|if only the positions of instrument
pointers|a fact whose relevance has been stressed with great force by
Bell~\cite{Bell82}.


\begin{thebibliography}{28.}


\bibitem{BG03} Bassi, A., Ghirardi, G.C.: ``Dynamical Reduction
  Models,'' Phys. Rep. \textbf{379}, 257--427 (2003),
  quant-ph/0302164.

\bibitem{BellBeables} Bell, J.~S.: ``Beables for quantum field
   theory,'' Phys.\ Rep.\ \textbf{137}, 49--54 (1986). Reprinted in
   Bell, J.~S.: \textit{Speakable and unspeakable in quantum mechanics}.
   Cambridge: Cambridge University Press (1987), p.~173.

\bibitem{Bell82} Bell, J.~S.: ``On the {I}mpossible {P}ilot {W}ave,''
   Found.\ Phys.\ \textbf{12}, 989--999 (1982).  Reprinted in Bell,
   J.~S.: \textit{Speakable and unspeakable in quantum mechanics}.
   Cambridge: Cambridge University Press (1987), p.~159.

\bibitem{Bohm52} Bohm, D.: ``A Suggested Interpretation of the Quantum
   Theory in Terms of ``Hidden'' Variables. I,'' Phys.\ Rev.\
   \textbf{85}, 166--179 (1952).  Bohm, D.: ``A Suggested Interpretation
   of the Quantum Theory in Terms of ``Hidden'' Variables. II,'' Phys.\
   Rev.\ \textbf{85}, 180--193 (1952).

\bibitem{BSB99} Brown, H., Sj\"oqvist, E., Bacciagaluppi, G.: 
  ``Remarks on identical particles in de Broglie--Bohm theory,''
  Phys. Lett. A \textbf{251}, 229--235 (1999), quant-ph/9811054.

\bibitem{identical} D\"urr, D., Goldstein, S., Taylor, J., Tumulka,
   R., Zangh{\`\i}, N.: ``Bosons, Fermions, and the Topology of
   Configuration Space,'' to be submitted.

\bibitem{density} D\"urr, D., Goldstein, S., Tumulka, R., Zangh\`\i,
   N.: ``On the Role of Density Matrices in Bohmian Mechanics,''
   to appear in Found.\ Phys.,
   quant-ph/0311127.

\bibitem{DGZ} D\"urr, D., Goldstein, S., Zangh\`\i, N.: ``Quantum
   Equilibrium and the Origin of Absolute Uncertainty,'' J.\ Stat.\
   Phys.\ \textbf{67}, 843--907 (1992), quant-ph/0308039.

\bibitem{op} D\"urr, D., Goldstein, S., Zangh\`\i, N.: ``Quantum
   Equilibrium and the Role of Operators as Observables in Quantum
   Theory,'' J.\ Stat.\ Phys.\ \textbf{116}, 959--1055 (2004),
   quant-ph/0308038.

\bibitem{Stanford} Goldstein, S.: ``Bohmian Mechanics'' (2001), in
   \textit{Stanford Encyclopedia of Philosophy}, edited by
   E.~N.~Zalta, published online by Stanford University,
   http://plato.stanford.edu/entries/qm-bohm.

\bibitem{conf} Goldstein, S., Taylor, J., Tumulka, R., Zangh{\`\i},
   N.: ``Are All Particles Real?'' quant-ph/0404134.

\bibitem{LDW71} Laidlaw, M.G., DeWitt, C.M.: ``Feynman functional
  integrals for systems of indistinguishable particles,''
  Phys. Rev. D \textbf{3}, 1375--1378 (1971).

\bibitem{Leinaas} Leinaas, J.~M., Myrheim, J.: ``On the Theory of
   Identical Particles,'' Il Nuovo Cimento \textbf{37}, 1--23 (1977).

\bibitem{stochmech} Nelson, E.: \textit{Quantum Fluctuations}.
   Princeton: Princeton University Press (1985).

\bibitem{Jamesthesis} Taylor, J.: ``Connections with Bohmian
  Mechanics,'' Ph.~D.\ Thesis, Department of Mathematics, Rutgers
  University (2003).

\bibitem{Val02} Valentini, A.: ``Subquantum Information and
  Computation,'' Pramana - J. Phys. \textbf{59}, 269--277 (2002), 
  quant-ph/0203049.


\end{thebibliography}
\end{document}